\documentclass[twocolumn,aps,prd,superscriptaddress]{revtex4-1}

\usepackage{graphicx}
\usepackage{gensymb}
\usepackage{dcolumn}
\usepackage{bm}
\usepackage{textcomp}
\usepackage[intlimits]{amsmath}
\usepackage{esint}
\usepackage{braket}
\usepackage{color}
\usepackage{filecontents}
\usepackage{bbold}
\usepackage{soul} 

\begin{document}
	
	
	\title{Angular memory effect of transmission eigenchannels}
	
	\author{Hasan Y{\i}lmaz}
	\affiliation{Department of Applied Physics, Yale University, New Haven, Connecticut 06520, USA}%
	\author{Chia Wei Hsu}
	\affiliation{Department of Applied Physics, Yale University, New Haven, Connecticut 06520, USA}%
	\affiliation{Ming Hsieh Department of Electrical and Computer Engineering, University of Southern California, Los Angeles, California 90089, USA}
	\author{Arthur Goetschy}
	\affiliation{ESPCI Paris, PSL University, CNRS, Institut Langevin, 1 rue Jussieu, F-75005 Paris, France}
	\author{Stefan Bittner}
	\affiliation{Department of Applied Physics, Yale University, New Haven, Connecticut 06520, USA}
	\author{Stefan Rotter}
	\affiliation{Institute for Theoretical Physics, Vienna University of Technology (TU Wien), 1040, Vienna, Austria}%
	\author{Alexey Yamilov}
	\affiliation{Department of Physics, Missouri University of Science \& Technology, Rolla, Missouri 65409, USA}%
	\author{Hui Cao}
	\email{hui.cao@yale.edu}
	\affiliation{Department of Applied Physics, Yale University, New Haven, Connecticut 06520, USA}%
	
	
	\begin{abstract}
		The optical memory effect has emerged as a powerful tool for imaging through multiple-scattering media; however, the finite angular range of the memory effect limits the field of view. Here, we demonstrate experimentally that selective coupling of incident light into a high-transmission channel increases the angular memory-effect range. This enhancement is attributed to the robustness of the high-transmission channels against perturbations such as sample tilt or wavefront tilt. Our work shows that the high-transmission channels provide an enhanced field of view for memory effect-based imaging through diffusive media.   
		
	\end{abstract}
	
	\pacs{Valid PACS appear here}

	
	\maketitle
	
	
	`Seeing through an opaque medium' has long been a grand challenge, as ballistic light decays exponentially with depth. Various techniques have been developed to extract the weak signal from single/few scattering in an overwhelming background of multiply-scattered light~\cite{2013_Park_OE, 2015_Choi_NatPhoton, 2016_Aubry_ScienceAdv, 2017_Wax_OptLett, 2019_Faccio_NatPhoton}. A recent paradigm shift is harnessing multiply-scattered or diffused light for imaging applications~\cite{2012_Mosk_NatPhoton_R, 2015_Yang_NatPhoton, 2015_Vellekoop_OptExpress, 2015_Choi_OptExpress_R, 2015_Park_R, 2017_Rotter_RMP_R}. The key ingredient that enabled this strategic shift is the hidden correlations of seemingly random speckles formed by the interference of scattered light~\cite{2012_Bertolotti_Nat, 2014_Fleischer_OE, 2014_Katz_NatPhoton, 2015_Yilmaz_Optica, 2018_Gehm_SR}. Quite remarkably, such correlations have been both predicted and observed in the angular, spectral, spatial, and temporal domains~\cite{1986_Shapiro_PRL, 1987_Genack_PRL, 1992_Lagendijk_PRB, 1994_BerkovitsPR, 2016_Mosk_OE, 2015_Gigan_OE, 2017_Vellekoop_Optica, 2018_Judkewitz_Optica, 1988_Feng_PRL, 1988_Stone_PRL, 1989_Berkovits_PRB, 1994_Genack_PRE, 2015_Yang_NatPhys}.
	
\begin{figure}[th]
	\centering
	\includegraphics[width=\linewidth]{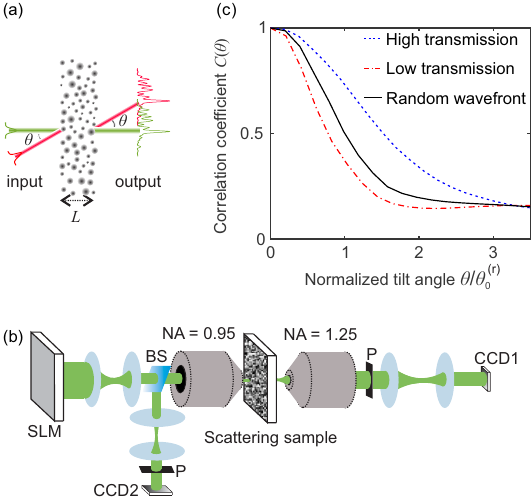}
	\caption{Angular memory effect, the experimental setup, and data. (a) Sketch of the angular memory effect for a diffusive slab. An incident beam generates a transmitted speckle pattern in far field (green). When the input wavefront is tilted by a small angle $\theta$, the output wavefront is tilted by the same angle $\theta$, leading to a lateral shift of far-field speckle pattern (red). (b) Simplified schematic of the experimental setup. The laser beam is modulated by a phase-only SLM, imaged onto the pupil of a microscope objective by a pair of lenses, and directed onto a ZnO nanoparticle film. The transmitted (reflected) light is measured by a CCD camera CCD1 (CCD2) in the far field. NA and P stand for numerical aperture and linear polarizer, respectively. (c) Experimentally measured intensity correlation function $C(\theta)$ of the transmitted speckle patterns as a function of the normalized tilt angle $\theta/ \theta_0^{\rm (r)}$, for a high-transmission channel (with $T/\overline{T} = 2.29$, blue dashed line), a low-transmission channel (with $T/\overline{T} = 0.58$, red dot-dash line), and a random incident wavefront (black solid line). $\theta_0^{\rm (r)}$ denotes the width of $C(\theta)$ for the random wavefronts where $C(\theta_0^{\rm (r)}) = C(0)/2$, and its value is about 1\textdegree.}
	\label{figure1}
	\end{figure}

	Perhaps the best known from all of the above correlations is the angular `memory effect': when the incident wavefront of a coherent beam on a diffusive medium is tilted by a small angle, the transmitted wavefront is tilted by the same amount, resulting in the translation of the far-field speckle pattern~\cite{1988_Feng_PRL, 1988_Stone_PRL, 1989_Berkovits_PRB, 1994_Genack_PRE} (see Fig.~\ref{figure1}(a)). The angular memory effect originates from the intrinsic correlations in the transmission matrix $t$ of a diffusive slab with a width $W$ that is much larger than its length $L$~\cite{1989_Berkovits_PRB, 1994_Genack_PRE, 2015_Yang_NatPhys}. In real space, $t$ is a banded matrix, because a point excitation at the front surface emerges as a diffuse halo of radius $L$ at the back surface of the slab. In the spatial-frequency domain, $t$ displays correlations between the matrix elements along the diagonal. The diagonal correlations are the origin of the memory effect with an angular correlation width $\theta_0 = \lambda/(2\pi L)$, where $\lambda$ is the wavelength of light. While the memory effect has already enabled various applications in imaging~\cite{2012_Bertolotti_Nat, 2014_Fleischer_OE, 2014_Katz_NatPhoton, 2015_Yilmaz_Optica, 2018_Gehm_SR}, its limited angular correlation width remains a central obstacle for wide-field imaging.
	
	A recent breakthrough in coherent control of light in diffusive media is the selective excitation of transmission eigenchannels by wavefront shaping~\cite{2007_Vellekoop, 2008_Mosk_PRL, 2012_Choi_NatPhoton, 2014_Popoff_PRL, 2016_Cao_PRL}. It allows not only to vary the transmittance from near zero to the order of unity, but also to drastically  change the spatial distribution of energy density inside the medium~\cite{2011_Choi_PRB, 2014_Aubry_PRL, 2015_Genack_NatCommun, 2016_Vos_NJP, 2016_Cao_PRL, 2018_Hong_Optica}. Moreover, it has very recently been discovered that in a wide diffusive slab, the transmission eigenchannels are localized in the transverse directions and have the same transverse width at the front and the back surfaces of the slab (i.e. there is no transverse spreading)~\cite{2019_Yilmaz_Nat_Photon, 2019_Tian_PRB}. Since the transverse spreading of scattered waves is inherently connected to the theory behind the angular memory effect, the absence of spreading immediately raises the question whether and how the angular memory effect is modified for transmission eigenchannels and whether one could make use of such modifications to increase the angular memory-effect range.
	
	In this Letter, we investigate this question experimentally and numerically by studying the angular memory effect of transmission eigenchannels in wide diffusive slabs. Compared to random incident wavefronts, we find that the angular memory-effect range is enhanced for high-transmission channels, but reduced for low-transmission channels. These phenomena can be explained by the robustness of the transmission eigenchannels against sample tilt or incident wavefront tilt. Our work illustrates the significance of high-transmission channels in memory-effect-based imaging applications: they not only penetrate deeper inside a diffusive medium, but also provide a wider field of view due to their enhanced angular memory-effect range. Furthermore, we observe the opposite behavior in reflection, where the angular memory-effect range is reduced for high-transmission channels but enhanced for low-transmission channels. This result suggests that the angular memory-effect range of reflected light may be used as a signature of coupling light into high-transmission channels in experiments where there is no access to the light field behind scattering media~\cite{2015_Park_OptCommun, 2015_Choi_OptExpress, 2018_Choi_NatPhoton}.
	
	Experimentally we measure the angular memory effect by selectively coupling coherent light into a single transmission eigenchannel. The scattering sample consists of densely-packed zinc oxide (ZnO) nanoparticles spin-coated on a cover slip. The thickness of the ZnO layer $L \simeq 10$ \textmu m is much smaller than its transverse dimensions (2 cm $\times$ 2 cm). Since the transport mean free path $l_{\rm t} \simeq 1.5$ \textmu m is much shorter than $L$, light transport in the sample is diffusive. The measured transmittance averaged over random incident wavefronts is $\overline{T} \simeq 0.2$.
	
	To find the transmission eigenchannels, we measure the field transmission matrix $t$ with the setup shown in Fig.~\ref{figure1}(b)~\cite{SM}. A monochromatic laser beam of wavelength $\lambda = 532$ nm is modulated by a phase-only spatial light modulator (SLM) before impinging on the sample. The transmitted field is measured by common-path interferometry with a CCD camera ~\cite{2010_Popoff_PRL, 2017_Hsu, 2019_Yilmaz_Nat_Photon}. We modulate two orthogonal polarizations of the incident field and record one linear polarization of the transmitted light. The field transmission matrix is obtained in spatial-frequency space. The incident wavefront $V_n$ of a transmission eigenchannel is determined from $t^\dagger tV_n = \tau_n V_n$, where $\tau_n$ is the $n$-th transmission eigenvalue (ordered from high to low $\tau_n$). We display the phase-front of $V_n$ on the SLM, and record the far-field intensity pattern of the transmitted field $I{(0)}$ with the CCD camera. 
	We then tilt an eigenchannel wavefront incident onto the sample by angle $\theta$ and track the change in the transmitted wavefront. The transmitted intensity pattern $I{(\theta)}$ on the camera is numerically tilted back by $\theta$, and its Pearson correlation with the original pattern $I{(0)}$ is computed as $C(\theta) = \langle \delta I{(0)} \delta I{(\theta)}\rangle/(\langle \delta I{(0)}^2\rangle^{1/2}\langle \delta I{(\theta)^2}\rangle^{1/2}) $, where $\delta I \equiv I - \langle I\rangle$ and $\langle ...\rangle$ represents spatial averaging over the output pattern. We calculate the intensity correlation coefficient $C(\theta)$ of the ten highest transmission channels, of the ten lowest transmission channels, and of twenty random incident wavefronts. In Fig.~\ref{figure1}(c) we show examples of $C(\theta)$ for a high and a low-transmission channel compared to that of a random wavefront: the high-transmission channel decorrelates slower with tilt angle $\theta$ than the random wavefront, while the low-transmission channel decorrelates faster. $C(\theta)$ does not decay to 0 at large $\theta$ due to the limited modulation efficiency of our SLM: as we tilt the incident wavefront with the SLM, a small portion of the field remains unmodulated, therefore the corresponding transmitted fields are correlated. From the width of $C(\theta)$, we determine that the angular memory-effect range for the highest transmission channel $\theta_0^{\rm (h)}$ is 1.52 times of that for a random wavefront $\theta_0^{\rm (r)}$, and the angular range for the lowest transmission channel is $\theta_0^{\rm (l)} = 0.77~\theta_0^{\rm (r)}$.
	
\begin{figure}[th]
    \centering
	\includegraphics[width=\linewidth]{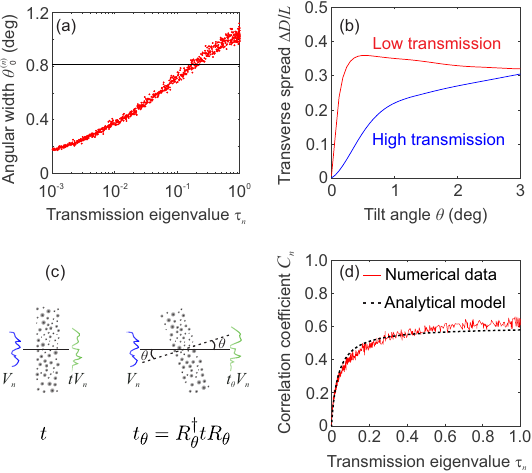}
	\caption{Numerical and theoretical results. (a) Angular correlation width $\theta_0^{(n)}$ of transmission eigenchannels versus their transmittance $\tau_n$. Each point represents an average over 10 disorder realizations. The horizontal black line denotes the angular correlation width for random incident wavefronts $\theta_0^{(r)}$. (b) Transverse spread $\Delta D$ of high ($\overline{\tau} < \tau_n < 1$, blue line) and low ($10^{-4} < \tau_n <\overline{\tau}$, red line) transmission eigenchannels vs. tilt angle $\theta$ of their incident wavefronts. (c) Sketch of the transmission matrix of a tilted sample $t_\theta = R_\theta^\dagger tR_\theta$, where  $R_{\theta}$ is the tilting matrix and $t$ the field transmission matrix without tilting. (d) Intensity correlation coefficient $C_n$ of transmission eigenchannels at the tilt angle $\theta = \theta_0^{\rm (r)} \approx 0.8$\textdegree, obtained from numerical simulation (red solid line), and the prediction of the phenomenological model, Eq.~(\ref{eq_01}), with $\sigma^2$ as the only fitting parameter (black dashed line). The number of channels is $N = 3239$. The diffusive slabs have thickness $k_0L = 100$, width $k_0W = 6000$, transport mean free path $k l_{\rm t} = 4.6$, average refractive index $n_0 = 1.5$, where $k = n_0k_0$, $k_0 = 2 \pi/ \lambda$, and $\lambda$ is the vacuum wavelength.}
	\label{figure2}
	\end{figure}

    To confirm that angular memory effect is enhanced for high-transmission channels and suppressed for low-transmission channels, we numerically simulate light propagation through two-dimensional (2D) diffusive slabs ($W \gg L \gg l_{\rm t}$). We calculate the complete field transmission matrix $t$ using the recursive Green's function method \cite{SM}. Evaluating the transmission eigenchannels of $t$, we calculate the output fields of each eigenchannel with respect to the tilt angle $\theta$ of its incident wavefront. The transmitted field is then tilted back by the same angle $\theta$, and its Pearson correlation with the original transmitted field is computed. From the field correlation $C^{\rm (E)}_n(\theta)$, the intensity correlation $C_n(\theta) = |C^{\rm (E)}_n(\theta)|^2$ is obtained. $C_n(\theta)$ decays with the tilt angle $\theta$, and its width $\theta_0^{(n)}$ gives the angular memory-effect range for the $n$-th eigenchannel. Fig.~\ref{figure2}(a) clearly shows that $ \theta_0^{(n)}$ increases with the transmission eigenvalue $\tau_n$. The eigenchannels with transmittance $\tau_n$ above the average value $ \overline{\tau}$ have larger memory-effect range, while those of $\tau_n <  \overline{\tau}$ have smaller memory-effect range than the random wavefronts. Furthermore, we find that the width $\theta_0^{\rm (h)}$ for high transmission channels is inversely proportional to the effective sample thickness $L_{\rm eff}$ that includes the extrapolation lengths~\cite{SM}.

	The numerically observed dependence of the eigenchannel angular memory effect on transmittance agrees qualitatively with the experimental observation. Such a dependence might be surprising at first sight as none of the eigenchannels of the \textit{complete} transmission matrix spreads laterally in the slab, and they all have the same transverse widths at the front and the back sides of the slab~\cite{2019_Yilmaz_Nat_Photon}. However, we should recognize that once the incident wavefront of an engenchannel is tilted, it is no longer the eigenvector of $t^{\dagger} t$. Consequently, lateral spreading occurs inside the slab, and the transmitted beam becomes wider than the incident beam for all eigenchannels. The effective widths of input and output beams are given by the participation numbers of the field intensity profiles at the front and the back surfaces of the slab~\cite{2019_Yilmaz_Nat_Photon}. Their difference $\Delta D$ is the transverse spread. As shown in Fig.~\ref{figure2}(b), $\Delta D$ increases as the tilt angle $\theta$ increases. However, the increase is much slower for high-transmission eigenchannels, indicating they are more robust against the tilt of the incident wavefront than the low-transmission eigenchannels. This leads to a larger memory-effect range for high-transmission channels than low-transmission ones. 
	
	The transmission eigenvalue dependence of the tilt-induced lateral spreading shown in Fig.~\ref{figure2}(b) can be understood as follows. When the incident wavefront of a transmission eigenchannel is tilted by an angle $\theta$, it excites not only this eigenchannel, but also other eigenchannels. The latter can be approximated as a random superposition of all other eigenchannels, which is equivalent to a random incident wavefront. The transmitted field profile at the back side of the sample is then a superposition of a transversely-localized eigenchannel profile and a transversely-spread random wavefront profile. For a high-transmission eigenchannel, its output profile dominates over the random wavefront profile, but for a low-transmission channel, the output is dominated by the random wavefront profile. As a result, the high-transmission channels have stronger correlation and larger memory effect than the low-transmission channels.
	
	To make the above understanding more quantitative, we introduce a phenomenological model that predicts $C_n$. As illustrated in Fig.~\ref{figure2}(c), the angular memory effect can be equivalently considered as the correlation of transmitted fields with respect to the tilt angle $\theta$ of the scattering sample for a fixed incident field. The transmission matrix of the tilted sample is $t_{\theta} = R_{\theta}^{\dagger}tR_{\theta}$, where the tilting matrix is written in the form $R_{\theta} = \mathbb{1} + X$. We model the matrix $X$ as an $N\times N$ complex random matrix with idependent and Gaussian-distributed entries. The variance $\sigma^2/N$ of its elements determines the amount of perturbation. When the incident wavefront corresponds to the transmission eigenchannel $V_n$ of the untilted sample, the transmitted field through the tilted sample is $t_{\theta}V_n$, and its correlation with the original transmitted field $ t \, V_n$ is	
	\begin{equation}
	\begin{split}
	\label{eq_01}
	C_n & \equiv \frac{| \braket{V_n|t^{\dagger}t_{\theta} |V_n} |^2} {\braket{V_n|t^{\dagger} t |V_n} \braket{V_n|t_{\theta}^{\dagger}t_{\theta} |V_n}} \\
	& \simeq \frac{1}{1 + \sigma^2}\frac{\tau_n + \sigma^4 {\overline{\tau}/{N}}} {\tau_n + \sigma^2\overline{\tau}},
	\end{split}
	\end{equation}
	(see~\cite{SM} for the derivation). The prediction of the model fits well to the numerical result (see Fig.~\ref{figure2}(d)), with a value of $\sigma$ that depends only on $\theta$ and the effective sample thickness $L_{\rm eff}$. In the limit $\theta\ll 1$ rad, it can be shown that  $\sigma \propto k_0L_{\rm eff}\theta$~\cite{SM}. Eq.~\eqref{eq_01} shows that the perturbed output is more correlated with the original output for high-transmission channels, and when $\tau_n \gg \sigma^2\overline{\tau}$, $C_n$ is on the order of unity. At the same time, the transmitted pattern decorrelates more for low-transmission channels. For $\tau_n\to 0$, $C_n$ is on the order of $1/N$, which is the expected value between two uncorrelated speckle patterns with $N$ speckle grains.
	
	\begin{figure*}[th]
	\centering
	\includegraphics[width=12.5cm]{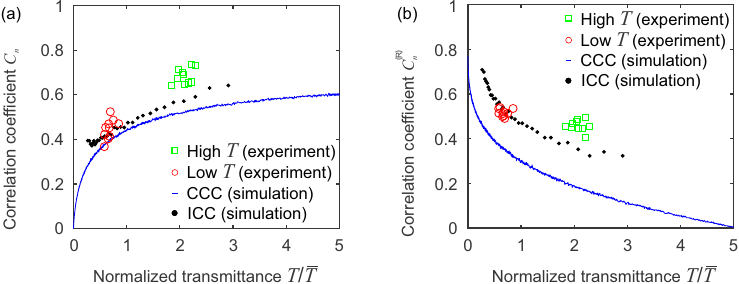}
	\caption{Comparison between the experimental and the theoretical results.  Correlation coefficient $C_n$ in (a) ($C_n^{\rm (R)}$ in (b)) of transmitted (reflected) eigenchannel intensity patterns for $\theta = 0$ and $\theta  = \theta_0^{\rm (r)}$ as a function of their normalized transmittance $T/\overline{T}$. $\theta_0^{\rm (r)}$ is the width at half maximum of the angular correlation function in transmission $C(\theta)$ for (a) and in reflection $C^{\rm (R)}(\theta)$ for (b) for random incident wavefronts. Blue solid line and black dots: numerical simulation results in case of complete channel control (CCC) and incomplete channel control (ICC), respectively. Green open squares (red open circles): experimental data for ten highest (lowest) transmission channels. The simulation data represents an average over 50 disorder realizations. The simulation parameters are slab thickness $L = 10$ \textmu m, width $W = 508$ \textmu m, $l_{\rm t} = 1$ \textmu m, $n_0 = 1.4$, background refractive index $n_1 = 1.0$ (in front of the slab), and  $n_2 = 1.5$ (at the back).}
	\label{figure3}
	\end{figure*}

	For a fair comparison of the simulation and the experimental data, we must take into account that only a limited number of channels is controlled in the experiment~\cite{SM}. The limited numerical aperture (NA) in the illumination and the detection, the finite area of illumination on the sample, the phase-only modulation of the (far-field) incident wavefronts, and single-polarization detection of the transmitted light all reduce the range of transmittance of experimentally realized eigenchannels~\cite{2013_Stone_PRL, 2014_Popoff_PRL, 2017_Hsu}. Such incomplete control also limits the enhancement or suppression of the angular memory-effect range that can be observed experimentally. Fig.~\ref{figure3}(a) shows the numerically calculated and the experimentally measured intensity correlation coefficient $C_n$ of transmission eigenchannels versus their normalized transmittance $T/\overline{T}$. The incomplete control reduces the ranges of both $C_n$ and $T/\overline{T}$. Despite the reduced range, the modification of the angular memory effect is clearly observed experimentally and agrees with the simulation result.
	
	A compelling question is raised by the enhanced memory-effect range for high-transmission channels: will the angular memory-effect range also be modified in reflection once light is coupled into a high-transmission channel? To answer this question, we experimentally measure the reflection correlations for individual transmission eigenchannels. The intensity pattern of reflected light is recorded in the far field by a second CCD camera (CCD2) in Fig.~\ref{figure1}(b). The modification of the angular correlations in reflection is opposite to the modification in transmission: the high-transmission channels' correlation is smaller in reflection than the low-transmission channels' correlation for a fixed tilt angle $\theta_0^{\rm (r)}$ (see Fig.~\ref{figure3}(b)). The angular correlation width $\theta_0$ in reflection for the highest (lowest) transmission eigenchannel is 7\% smaller (6\% larger) than that for the random incident wavefronts. Our numerical simulation confirms the experimental observation: the intensity correlation coefficient in reflection $C_n^{\rm (R)}$ decreases as the transmittance increases (Fig.~\ref{figure3}(b)). Taking into account the incomplete control in our experiment, the numerical results are in good agreement with the experimental data. 
	
	The modification of the angular memory effect in reflection can also be explained in the framework of our phenomenological model by replacing $\tau_n$ by $1 - \tau_n$~\cite{SM}. Once the incident light is coupled into a high (low) transmission channel, the reflectance is low (high) and the reflected field pattern is sensitive (robust) to the sample tilt. The reduced memory-effect range in reflection may provide experimental guidance for shaping the incident wavefront to couple light into high-transmission channels when there is no access to the transmitted light~\cite{2015_Park_OptCommun, 2015_Choi_OptExpress, 2018_Choi_NatPhoton}.
	
	In summary, we demonstrate that the angular memory effect for individual transmission eigenchannels is distinct from that of random wavefronts. With increasing transmittance, the eigenchannel memory-effect range increases in transmission, but decreases in reflection. Such variations can be explained by our phenomenological model in terms of the robustness of the eigenchannels against perturbations such as a sample tilt or an incident wavefront tilt. Our model can be extended to other perturbations, such as frequency detuning of the incident light, and provides an understanding of the enhanced bandwidth (spectral memory effect) for high-transmission channels, which was observed previously~\cite{2016_Mosk_OE}. Therefore our work reveals the general characteristic of high-transmission channels: their transmitted fields are robust while their reflected fields are sensitive against perturbations. Thanks to their larger angular memory-effect range, the high-transmission channels provide a wider scan range than Gaussian beams or random wavefronts, which will be useful for improving the quality of memory-effect-based speckle imaging through diffusive or otherwise complex media. Finally, the spatial memory effect was recently discovered in anisotropic scattering systems of length much larger than the scattering mean free path but comparable to or smaller than the transport mean free path \cite{2015_Yang_NatPhys, 2017_Vellekoop_Optica, 2018_Judkewitz_Optica}. It will be interesting to investigate the spatial memory effect for the transmission eigenchannels of such systems. 
	
	\begin{acknowledgements}
	We thank Allard Mosk (Utrecht University) for his comments after critically reading the manuscript. This work is supported partly by the Office of Naval Research (ONR) under grant no. MURI N00014-13-0649 and by the US-Israel Binational Science Foundation (BSF) under grant no. 2015509. We acknowledge support from the European Commission under project NHQWAVE (Grant No. MSCA-RISE 691209), as well as the computational resources by the Yale High Performance Computing Cluster (Yale HPC).
	\end{acknowledgements}
	
	\bibliography{eigenchannel_memory_effect}
	
\end{document}